\documentclass[fleqn,twoside]{article}
\usepackage{espcrc2}
\usepackage{psfig}

\usepackage{graphicx}
\usepackage[figuresright]{rotating}

\newcommand{\AmS}{{\protect\the\textfont2
  A\kern-.1667em\lower.5ex\hbox{M}\kern-.125emS}}

\newcommand{\aap}{A\&A}

\newcommand{\aj}{AJ}
\newcommand{\mnras}{MNRAS}

\newcommand{\ltsimeq}{\raisebox{-0.6ex}{$\,\stackrel 
        {\raisebox{-.2ex}{$\textstyle <$}}{\sim}\,$}} 
\newcommand{\gtsimeq}{\raisebox{-0.6ex}{$\,\stackrel 
        {\raisebox{-.2ex}{$\textstyle >$}}{\sim}\,$}}

\title{Galaxy Evolution and Cosmology with the Square Kilometre Array}

\author{
S. Rawlings\address[Ox]{Astrophysics, Department of Physics, Keble Road, Oxford, OX1 3RH, UK},
F.B. Abdalla\addressmark[Ox],
S.L. Bridle\address{Department of Physics and Astronomy, University College London, London, WC1E 6BT, UK},
C.A. Blake\address{School of Physics, University of New South Wales, Sydney, NSW 2052, Australia},
C.M. Baugh\address[ICC]{Institute for Computational Cosmology, 
        Department of Physics, 
        University of Durham, South Road, Durham, DH1 3LE, UK}
L.J. Greenhill\address{Harvard-Smithsonian Center for Astrophysics, 60 Garden St, Cambridge,
MA 02138, USA},\\
J.M. van der Hulst\address{Kapteyn Institute, Postbus 800, NL-9700 AV Groningen, the 
Netherlands}
}

\begin{document}

\begin{abstract}
The present-day Universe is seemingly dominated by dark energy and dark matter, but
mapping the normal (baryonic) content remains vital for both astrophysics -- understanding how 
galaxies form -- and astro-particle physics -- inferring properties of the dark components.
 
The Square Kilometre Array (SKA)
will provide the only means of studying the cosmic evolution of neutral Hydrogen (HI) 
which, alongside information on star formation from the radio continuum, is
needed to understand how stars formed from gas within
dark-matter over-densities and the r\^{o}les of gas accretion and galaxy merging.

 `All hemisphere' HI redshift surveys to $z \sim 1.5$ are feasible with 
wide-field-of-view realizations of the SKA and, by measuring the galaxy power spectrum in 
exquisite detail, will allow the first precise studies of the equation-of-state of dark energy.
The SKA will be capable of other uniquely powerful cosmological studies including the measurement 
of the dark-matter power spectrum using weak gravitational lensing, and the precise
measurement of $H_{0}$ using extragalactic water masers.

The  SKA is likely to become the premier 
dark-energy-measuring machine, bringing breakthroughs in cosmology beyond those likely to be
made possible by combining
CMB (e.g.\ Planck), optical (e.g.\ LSST, SNAP) and other early-21$^{\rm st}$-century datasets. 
\end{abstract}

\maketitle

\section{Introduction}
\label{intro}

As detailed by Blake et al.\ (2004; hereafter BABR), the
`era of precision cosmology' has been ushered in by the measurement of many of the 
key cosmological parameters. We can now seemingly
adopt a fiducial model consisting of a spatially-flat ($\Omega_k = 0$)
Universe with cosmological parameters $h \simeq 0.7$, $\Omega_{\rm m} \simeq
0.3$, $\Omega_{\rm b}/\Omega_{\rm m} \simeq 0.15$, $\sigma_8 \simeq 1$ and
$n_{\rm scalar} \simeq 1$ (values broadly consistent with the most recent CMB results,
Spergel et al.\ 2003). The big surprise of this `concordance cosmology'
has been the compelling evidence for a smoothly distributed
dark energy (with a density parameter $\Omega_{\rm DE} \simeq 0.7$).
The dark energy model is commonly characterized by its {\it equation-of-state}
$w(z) = P/\rho$ relating its pressure $P$ to its energy density $\rho$
(in units where the speed-of-light $c = 1$). 
For consistency with BABR we adopt a fiducial dark energy model
characterized by $w(z) = w_0 + w_1 z$ where
$(w_0,w_1) = (-0.9,0)$.\footnotemark

\footnotetext{
This has been chosen to illustrate the accuracy with which
the SKA will be able to discriminate a general dark energy model from one in which the 
dark energy is Einstein's cosmological constant $\Lambda$ for which $(w_0,w_1) = (-1,0)$.  
}

Results from a combination of early-21$^{\rm st}$-century
experiments will be essential in transitioning from the
10-per-cent accuracy currently achieved by current `precision' cosmology to the 
1-per-cent accuracy needed, for example, to make the first serious study of the properties of dark energy.
This leap in precision will be essential in making progress towards answering 
at least four of the eleven questions raised in {\em ``Connecting Quarks with the Cosmos: Eleven Science
Questions for the New Century''} (ref.\ 10).

Alongside this revolution in cosmology there has also been great progress in understanding
how galaxies and larger-scale structures form. Galaxies are visible because of their
normal (baryonic) content, and our understanding of the complex physics describing the r\^{o}le of baryonic
material in galaxy formation is growing but incomplete. 
`Semi-analytic' models (e.g.\ Cole et al.\ 2000; Baugh et al.\ 2004a, hereafter B2004) 
have become the primary mode of investigating these complex issues. B2004 outline one recipe 
(GALFORM) for galaxy formation, comprising nine physical ingredients. A measure of the
uncertainties involved is that just one of these ingredients, the hierarchical growth of dark-matter
halos, is relatively uncontroversial. However, for example, even ingredient (ii) from the Baugh et
al.\ recipe -- the virialization by shocks of gas within gravitational potential wells -- is now the subject of 
intense debate (e.g.\ Binney 2004; Keres et al.\ 2004).

A suite of surveys for neutral Hydrogen (HI) emission will be possible with the SKA
(van der Hulst et al. 2004, hereafter V2004).
In Sec.~\ref{sec:surveys} we discuss the number of galaxies 
expected in these surveys, and the uncertainties involved. In Sec.~\ref{sec:gformation} we
explain how measurements of HI will tell us new 
things about how galaxies form and evolve. We also briefly discuss 
ways other than HI surveys in which the SKA will improve our physical understanding of galaxy formation.
In Sec.~\ref{sec:cosmology} we show that, regardless of differences between 
predictions and the eventual measurements, moderate resolution ($\sim 1$ arcsec)
HI surveys with the SKA will help 
revolutionize cosmology providing the SKA realization has 
sufficient instantaneous field of view. We emphasise the importance of two other
cosmological experiments with the SKA: (i) a weak gravitational lensing survey, also
requiring large field of view, but higher ($\sim 0.1$ arcsec)
spatial resolution; and (ii) accurate measurement of the
Hubble constant using water masers, requiring high frequency and very-high-resolution capabilities.
We briefly contrast the 
impact of the SKA on measurements of dark energy with those of other planned 
early-21$^{\rm st}$-century experiments.

\section{HI surveys with the SKA}
\label{sec:surveys}

Neutral atomic Hydrogen (HI) is the most abundant element in the Universe, the main
fuel for star formation, and a very powerful tracer of galaxies in various stages of
their evolution and in a variety of environments. The best way to detect the presence of HI is by its 21-cm
emission line, so {\it only} radio telescopes are capable of imaging the
distribution and kinematics of HI in the Universe. Although telescopes at other 
wavelengths will continue to probe the history of star formation, we
will never have a complete picture until we probe the the state of
the component from which the stars formed -- the neutral Hydrogen.
The SKA will be the ultimate HI imaging machine which will map out where 
the HI resides over a large range in cosmic time and galactic 
environments, from clusters to filaments to voids.

\begin{figure}[htb]
\includegraphics[width=7.5cm]{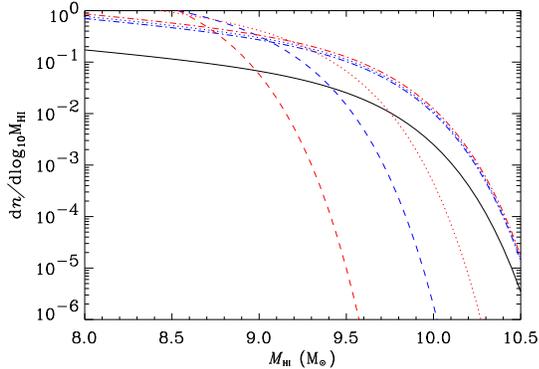} 
\vspace*{-0.5in}
\caption{\small 
\label{fig:h1lfa}
Predictions of the evolution in the HI mass function from AR2004.
The dot-dashed lines show their `Model A' at $z = 1$ (lower, blue) and $z = 2$ (upper, red) with
the solid (black) line showing the measured local HI mass function (Zwaan et al.\ 2003). 
The dashed lines show their `Model B' at $z = 1$ (rightmost, blue) and $z = 2$
(leftmost, red), and the dotted lines their `Model C' at $z = 1$ (upper, blue) 
and $z = 2$ (lower, red).}
\end{figure}

\begin{figure}[htb]
\center
\includegraphics[width=7.5cm]{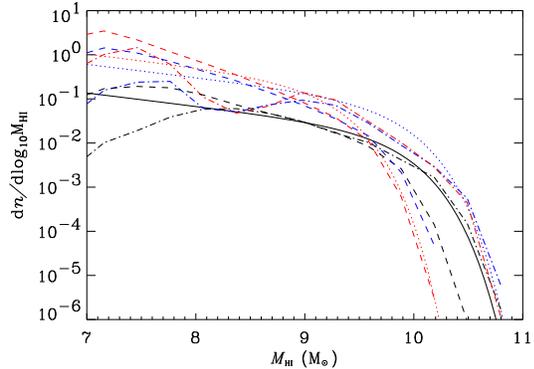} 
\vspace*{-0.5in}
\caption{\small 
\label{fig:h1lfb}
Comparison of HI mass functions from
Model C (dotted, AR2004), a `no evolution' model (solid, black) with
those from the two versions of GALFORM described by B2004: 
dot-dashed, the `superwind' model (Benson et al.\ 2003); and dashed, the `standard' semi-analytic model
of Cole et al.\ (2000). Lines at $z = 0$ (black), $z = 1$ (blue) and $z = 2$ (red)
are shown for the semi-analytic models; for the superwind model the $z=1$ and $z=2$ curves
are practically indistinguishable, both lying above the $z=0$ curve; for the standard model, the
$z=0,1,2$ curves are progressively to the left of the no-evolution curve at high HI masses.}
\end{figure}

One very-wide-area (`all hemisphere') survey of HI has already been performed
(the HIPASS survey; Meyer et al.\ 2004) but because of its limited sensitivity it reaches
only to redshift $z \approx 0.04$. Although this survey has demonstrated that most HI is associated 
with galaxies, we do not yet have any meaningful
measurement of the redshift distribution ${\rm d} N  / {\rm d} z$ of 
these HI-emitting galaxies, and we must rely on models. 
Abdalla \& Rawlings (2004; hereafter AR2004) investigated a 
range of evolutionary scenarios for HI-emitting galaxies, comparing a 
`no evolution' benchmark with models obeying the integral constraint coming from 
the inferred rise with redshift of the total density of HI (derived from the statistics
of HI absorption lines in the optical spectra of distant quasars; Peroux et al.\ 2001). 
In Fig.~\ref{fig:h1lfa} the range in evolutions in the HI mass function from AR2004
are shown and in Fig.~\ref{fig:h1lfb} their preferred Model `C' is
compared to those arising from two versions of GALFORM (B2004).

It is gratifying, if unsurprising, that at modest redshifts ($0 \leq z \leq 1$)
there is good agreement between the (preferred) `Model C' of AR2004 and the 
results of the `superwind' semi-analytic model (B2004). The
AR2004 Model C is implicitly hard-wired to reproduce the local HI mass function and
has an evolutionary behaviour which tracks `blue' (quiescently) star-forming
blue galaxies, many properties of which are now well established at $z \sim 1$ 
(see AR2004). The superwind semi-analytic model is effectively hard-wired to produce reasonable 
low-redshift HI results because its adjustable parameters have been 
set to values capable of reproducing the optical luminosity function of galaxies, and, for 
blue star-forming galaxies at least, the ratio of HI mass $M_{\rm HI}$ to blue luminosity is
fairly constant. The `standard' semi-analytic model fails to
predict the $z = 0$ HI mass function and, like Model B of AR2004, is disfavoured by the
properties of blue galaxies at $z \sim 1$ (AR2004); its evolutionary behaviour
is probably ruled out by its poor fit to modern datasets (B2004). The superwind model
fits available datasets well and will score a notable success 
if its prediction of the HI mass function at $z = 2$ 
proves to be in accord with data obtained with the SKA.

\begin{figure}[htb]
\includegraphics[width=7.5cm]{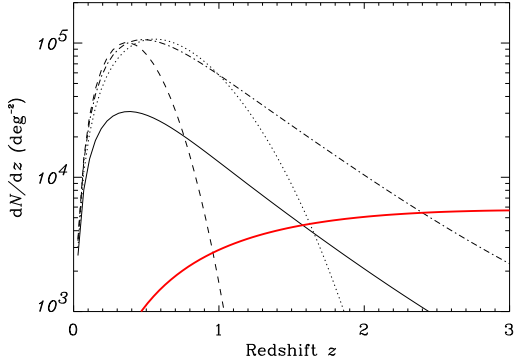} 
\vspace*{-0.5in}
\caption{\small 
\label{fig:dndz}
Predictions of ${\rm d} N  / {\rm d} z$ per deg$^{2}$ for an SKA survey with an
exposure time of 4 hours, a signal-to-noise detection limit of $10$ and assumptions about the
properties of the HI-emitting galaxies and the SKA detailed in Sec.~\ref{sec:surveys}. The same
linestyles are used as in Fig.~\ref{fig:h1lfa} to discriminate between the different
AR2004 models; the prediction of a `no-evolution' model is shown by the solid (black) line.
Also shown (thicker red line) is the surface density of galaxies needed for
a survey to be limited by cosmic variance rather than shot noise (AR2004).}
\end{figure}

Until such data are available, however, we show, in Fig.~\ref{fig:dndz}, the 
variation in the redshift distribution
${\rm d} N  / {\rm d} z$ predicted by the full set of models considered by AR2004. 
As well as obvious variations in the redshift distributions, there are factor 
$\sim 4$ variations in the total number of objects expected. However, in all cases, we can
be confident of large numbers\footnotemark ($\gtsimeq 10^{4} ~ \rm deg^{-2}$) of 
intermediate redshift ($0.5 < z < 1.5$) HI-emitting galaxies in an SKA survey of the depth 
considered.

\footnotetext{
This is vital for cosmological applications  
(Sec.~\ref{sec:cosmology}) where the uncertainties in the 
galaxy power spectrum need to be dominated by cosmic variance not shot noise 
(Fig.~\ref{fig:dndz}; AR2004).} 

Measurements of ${\rm d} N  / {\rm d} z$ with the 
SKA will obviously remove the uncertainties obvious in Fig.~\ref{fig:dndz} and will, together
with other SKA-based experiments, provide crucial new information on 
galaxy formation and evolution (Sec.~\ref{sec:gformation}). Deep-field observations with the
SKA will push these studies to very high redshift ($z \gtsimeq 3$; V2004).

Reproducing Equation~6 of AR2004, for a given time of survey per square degree,
the limiting HI mass that the SKA 
will be able to see is 

\begin{equation}
\label{eqn:mass}
M_{\rm HI}(z)= k \frac{D_L^2(z)}{(1+z)^{1+p}} 
\frac{\sqrt{V(z) {\Delta}V}}{f} S_{\rm N} \frac{\sigma_{\rm 8h}}{\sqrt{2}} 
\sqrt{\frac{8}{t}},
\end{equation}

\noindent 
where $k$ 
is a known\footnotemark constant (AR2004), $D_{\rm L} (z)$ is the luminosity distance,
\footnotetext{
Note that $k$ varies with the SKA realization. AR2004 argue that any realization will need to have 
a sizeable fraction ($\sim 50$ per cent) of its collecting area in a compact
(5-km diameter) core to avoid losing sensitivity to extended HI emission.
}
$p$ is defined by the field of view changing with frequency $\nu$
as $\nu^{-2p}$ (i.e.\ $p=1$ for the simple diffraction-limited case);
$f$ is the fraction of the sensitivity relative to the SKA, 
$V(z)$ is the line-of-sight width corresponding to the projected circular velocity of the galaxy
(assumed to evolve systematically with $z$),
$\Delta V$ is the SKA channel width, $S_{\rm N}$ is the signal-to-noise level of detections;
$\sigma_{\rm 8h}$ is the noise for an eight-hour integration (per $\Delta V = 30 ~ \rm km s^{-1}$ channel)
for each of two polarizations, its value $\sim 2 ~ \rm \mu Jy$ increasingly weakly with
declining frequency and hence increasing redshift); and $t$ is 
the integration time in hours for a given value of $FOV$.\footnotemark
\footnotetext{
We emphasise that $FOV$ is defined to be the usable field-of-view
at an observed frequency of 1.4 GHz.
}

By definition, $f=1$ for the SKA. Indeed, from Fig.~\ref{fig:dndz}, it is clear that this
value of $f$ is sufficient and necessary\footnotemark to detect (and measure redshifts for)
large numbers of HI galaxies out to $z \sim 1.5$ in a few hours of SKA exposure. 
For large-sky-area surveys, the value of $p$ is crucial. AR2004
consider the case where the pointing centres of a survey are `tiled' such that there is only a
small overlap in the fields imaged at good (and roughly uniform) sensitivity at 1.4 GHz. In 
SKA designs with $p > 0$, the sky area
over which surveys pick up HI-emitting objects increases with increasing redshift, and useful
data at (say) $z \sim 1$ is obtained from several different pointing positions.
SKA designs with large values of $FOV$ (and a $\nu^{-2}$ dependence) are therefore
far more efficient at covering sky area, and hence cosmic volume. AR2004 demonstrate that 
`all hemisphere' surveys become feasible provided,

\begin{equation}
\label{eqn:fov}
\beta \, \, FOV \, \, T_{0} \, \gtsimeq \, 10,
\end{equation}

\noindent
where $\beta$ is the ratio of the instantaneous SKA bandwidth to the bandwidth required by the 
survey, and $T_{0}$ is the total duration of the survey in years.
The case for the widest sky coverage possible is made in Sec.~\ref{sec:cosmology}, so
$FOV$ must be as large as possible and 
$\beta$ needs to allow for the coverage of the key $0.5-1.4$ GHz range in the fewest possible settings.

\footnotetext{
As the time to survey a given sky area to a given limiting flux density (the `mapping speed') scales as
$f^2 \, FOV$, achieving at least $f \sim 1$ is clearly crucial.
}

The properties of the HI-emitting galaxies also influence their detectability with the 
SKA. AR2004 consider these effects in detail (their assumptions have been 
adopted for Fig.~\ref{fig:dndz}), taking intrinsic disk sizes that scale $\propto (1+z)^{-2}$
and velocities $V$ that scale $\propto (1+z)^{-0.5}$.

\section{Galaxy evolution with the SKA}
\label{sec:gformation}

\subsection{Context}
\label{sec:context_gformation}

Although we eagerly await proof by direct detection of a dark-matter particle, 
there is now a broad consensus that, 
after dark energy, the second most important contributor to the present-day energy budget of 
the Universe is Cold Dark Matter (CDM). In the context of the $\Lambda$CDM model,
the gravitational collapse of dark-matter halos has proved a compelling model for the formation
of structure in the Universe. Robust numerical (`N-body') calculations are available for
$\Lambda$CDM (e.g.\ Jenkins et al.\ 2001) which 
are in quantitative agreement with conceptually simple analytic approximations based on the 
Press-Schechter formalism. The growth of structure is hierarchical with 
more massive structures forming, over time, from the merging of less massive precursors. 
The mass function of dark-matter halos can be calculated by simply
evolving the tiny fluctuations in the density field at recombination, as measured by CMB experiments at $z \sim 1100$,
to the present-day ($z=0$) Universe. 

\begin{figure}[htb]
\includegraphics[width=7.5cm]{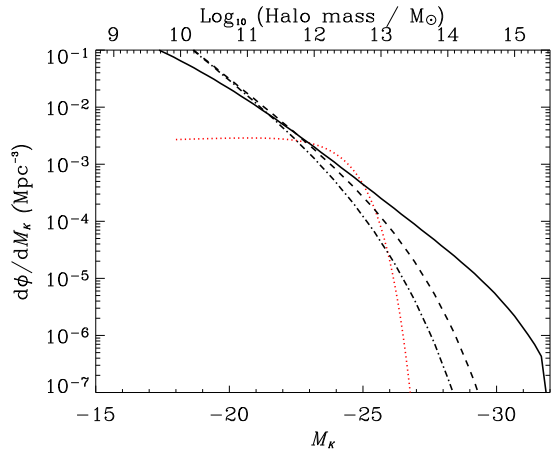} 
\vspace*{-0.5in}
\caption{\small 
\label{fig:ps}
Comparison of the measured $K-$band luminosity function of galaxies ${\rm d} \phi / {\rm d} M_{K}$,
where $M_{K}$ is the absolute magnitude, (approximated by the red, dotted Schechter function from
Cole et al.\ 2001), 
with the prediction of a toy model in which the $z = 0$ dark-matter mass function is converted to 
a luminosity function by assuming a constant mass-to-light ratio $\Gamma = 25 ~ \rm M_{\odot} / L_{K,odot}$
(solid black line calculated from the $\Lambda$CDM fitting function of 
Jenkins et al.\ 2001). Also shown are the dark-matter mass functions at $z=2$ (black dashed line)
and $z = 4$ (black dot-dashed line); note that 
straightforward interpretation of all the dark-matter mass functions requires use of the 
upper $x$-axis. This illustrates that the exponential decline in the 
Schechter function might be imprinted by one of more types of dramatic feedback events shutting 
off star formation in halos at high redshift (e.g.\ Rawlings \& Jarvis 2004 and refs. therein).}
\end{figure}

We show in Fig.~\ref{fig:ps} the near-infrared ($K-$band) luminosity function of galaxies which would be inferred 
from the mass function of dark-matter halos if we adopted a universal mass-to-light ratio $\Gamma$. We 
chose $\Gamma$ so that $L_{*}$ galaxies have roughly twice the space density of dark-matter halos of
mass $\sim 2 \times 10^{12} ~ \rm M_{\odot}$. This is line with the observation that a typical $L_{*}$
galaxy is grouped with $\sim 1$ other $L_{*}$ galaxy (Eke et al.\ 2004). 
The value adopted ($\Gamma = 25 ~ \rm M_{\odot} / L_{K,\odot}$) is also 
in line with direct measurements of mass-to-light ratios in systems dominated by
binary $L_{*}$ galaxies. It is obvious from Fig.~\ref{fig:ps} that a close mapping 
between the dark-matter mass function and the galaxy luminosity function occurs only at this
specific mass scale, with a clear lack of starlight in both more and less massive halos.
A complicated mapping is, of course, expected because 
galaxies are made up of baryonic material which, by radiating away or 
absorbing energy, or by participation in bulk flows, can
achieve very different spatial distributions to the CDM.

There have been many attempts to understand the complicated behaviour of baryons in the 
context of galaxy formation based, typically, on semi-analytic techniques
(e.g.\ Benson et al.\ 2003; B2004). The lack of low-mass galaxies 
is explained by baryons being heated to high enough temperatures that they have difficulty in becoming bound
to low-velocity-dispersion halos. It is not clear whether 
this heating is a `universal' process like reionization (in which case,
perhaps, the only low-mass galaxies to form were those which formed before reionization in protoclusters) or is the
result of more localised and recent feedback effects due, e.g., to supernovae-driven winds resulting from the
first bursts of star formation (Dekel \& Silk 1986). 
The lack of high-mass galaxies is linked to the very long cooling times
of the low-density gas which is associated with these halos (White \& Rees 1978), with the
low densities being the inevitable result of these halos collapsing only recently in an expanding Universe.
However, this process alone cannot explain the exponential decline in the number density of the most
massive galaxies (e.g.\ Benson et al.\, 2003), and additional 
physical feedback mechanisms are invoked. These range from the fairly weak 
(but common) feedback events associated with starburst-driven winds (e.g.\ Silk 2003), 
through quasar-driven winds (Silk \& Rees 1998) 
to ultra-powerful (but rare) bow shocks associated with radio galaxies, the latter carrying
sufficient power to influence hundreds of galaxies in protoclusters (Rawlings \& Jarvis 2004). 

With such a bewildering range of physical processes, and of course many adjustable 
parameters in their semi-analytic manifestation, there are great 
challenges in developing semi-analytic models
with any serious predictive power. Modellers are responding to these challenges by
extending, as increasing computer power allows, the use of direct numerical simulations of baryon-related physics,
and also by confronting their models with data collected across all the available wavebands (B2004; 
Baugh et al.\ 2004b).
Alongside these efforts there is a clear imperative to obtain the best datasets possible in the realistic
hope that models and data will eventually converge on a coherent description of galaxy formation and evolution.

\subsection{The SKA r\^{o}le}
\label{sec:gevolution}

We consider first HI surveys with the SKA (Sec.~\ref{sec:surveys}; AR2004, V2004). Assuming 
a wide-field-of-view solution for the SKA (Eqn.~\ref{eqn:fov}), an `all hemisphere' HI SKA
survey would contain $\sim 10^{9}$ galaxies out to $z \sim 1.5$ (AR2004). This redshift survey
would increase the cosmic volume, and also the number of $\sim L_{*}$ objects surveyed, by a factor $\sim 1000$
over current state-of-the-art optical surveys like the 2-degree-Field Galaxy Redshift Survey 
(2dFGRS) and the Sloan Digital Sky Survey (SDSS). The HI properties of more distant galaxies will be probed
by deep pointed SKA surveys (sometimes targeted on massive rich clusters to exploit the 
benefits of amplification by gravitational lensing). Although sensitivity considerations will limit
studies of HI emission to $z \sim 3$ (see Kanekar \& Briggs 2004 for methods of
probing higher redshift with HI absorption), if we, once again, assume a wide-field-of-view SKA
realization, very large samples of distant galaxies can be studied in HI.
A single-pointing SKA ultra-deep field totalling $\sim 30$ days exposure would
reach a 5-$\sigma$ HI mass limit $\sim 5 \times 10^{9} ~ \rm M_{\odot}$ (Fig.~1 of AR2004; V2004) and hence 
should (Sec.~\ref{sec:surveys}) get close to the break in the HI mass function at $z \sim 3$. Assuming
$FOV = 10 ~\rm deg^{2}$ and $p=1$, the survey would have an effective field of view of 
$\approx 160 ~ \rm deg^{2}$ and detect $\sim 10^{6-7}$ high-redshift 
($2.5 \leq z \leq 3.5$) galaxies. The surface density of these
objects would therefore be up to $\sim 10$-times higher than objects detected currently 
by the optical Lyman-break technique, but again the key
breakthrough would come by virtue of the huge field of view, generating a high-$z$ 
redshift survey outnumbering current samples by a factor $\gtsimeq 1000$. 

Regardless of the unique information provided by HI, statistical studies of galaxies in the range
$0 \ltsimeq z \ltsimeq 4$, will be revolutionised by the SKA. In Sec.~\ref{sec:cosmology} we focus on the
cosmological applications of measuring the galaxy power spectrum, but essentially all the
investigations made possible by the 2dFGRS and SDSS programmes
(i.e.\ the dependence of galaxy properties on internal and dark-matter halo properties, as well as
larger-scale clustering properties), could be pursued by the SKA across a much larger range
of redshifts, and with errors bars on many of the critical calculated quantities reduced 
by factors $\sim \sqrt{1000}$. This, together with weak-lensing measurements
of the dark matter (Sec.~\ref{sec:exp2}), will yield a detailed picture of
how galaxies, of all masses and in all environments, trace the underlying dark-matter
fluctuations. It has become clear from the 2dFGRS and SDSS results that this 
so-called `galaxy bias' is likely to be stochastic, scale-dependent, non-local and non-linear 
(e.g.\ Wild et al.\ 2004). Galaxy
surveys of the size generated by the SKA will be needed to obtain a full understanding of galaxy
bias which lies at the very heart of understanding how galaxies form and evolve.

To focus now on the new information provided by HI, we note first that the presence of a galaxy in 
an SKA redshift survey will be determined by the cold
gas mass ($M_{\rm gas}$) rather than the complicated mix of stellar mass
and star-formation rate familiar from, e.g.\, optical Lyman-limit surveys. 
It should therefore be possible to make much cleaner tests of specific models for galaxy formation and evolution.
For example, recent debate concerning galaxy evolution has highlighted 
uncertainties due to the variation of the timescale for star formation with epoch and halo (dark-matter) mass.
Different assumptions about this timescale 
(defined by $\tau_{*} = M_{\rm gas} / \psi$, where $\psi$ is the star formation rate) 
are the main reason for the huge difference between the predictions of the two semi-analytic models shown in
Fig.~\ref{fig:h1lfb} (B2004). SKA surveys will measure $M_{\rm gas}$ via HI
and estimate $\psi$ via non-thermal and thermal radio emission (with spatial and spectral separation of these
components available to refine these estimates; V2004). Gas and star formation 
properties can be studied as a function of dark-matter properties through measurement of internal 
(`Tully Fisher') HI velocity dispersions and cluster velocity dispersions.

In addition to the gas trapped in galaxy potential wells, where the star 
formation takes place, SKA will also chart the HI environments and
hence give a handle on the mechanism triggering star formation and the 
process of gas and/or satellite galaxy accretion. An obvious illustration of the 
diagnostic power of HI imaging is the nearby Andromeda group (Yun et al.\ 1994). 
Without the HI information, which clearly shows that the galaxies are ``communicating'', 
our interpretation of the star-formation activity in, e.g.\ M82, would not 
be as solid as it is now. 

SKA surveys should make clear how spiral galaxies formed and evolved. 
They will show where and when the gas is converted into stars and whether 
galaxy formation is dominated by gas accretion or the process of 
galaxy merging. The HI mass function
at low redshift is dominated by such late-type systems, and as discussed in Sec.~\ref{sec:surveys}, we have some
confidence in extrapolating their `quiescent' mode
of star formation at least as far as redshifts $z \sim 1$. However, 
as should be clear from Fig.~\ref{fig:h1lfb}, current models make very different predictions as to the 
precursors of these galaxies at even moderate redshifts $z \sim 2$. Indirect
evidence from HI absorption (Kanekar \& Briggs 2004; AR2004) suggests that, at
$z \sim 2$, most of these systems are not the well-formed disc systems we see today, but smaller
precursor objects. The SKA will directly map the way in which these
objects merge to form the local population of spiral galaxies.

SKA surveys should also map out how giant elliptical galaxies formed and evolved. A persistent 
debate in astrophysics has been whether these objects formed their stars during a `monolithic' collapse,
or were built up hierarchically.
This debate is sometimes couched as a fundamental test of $\Lambda$CDM models, but this
is rather misleading. From Fig.~\ref{fig:ps}, we can see that the addition of feedback mechanisms to
the standard hierarchical picture can explain the origin of the exponential decline in the 
galaxy luminosity function as the result of `monolithic' creation events associated with dramatic 
feedback episodes. This debate is once more intimately related to assumed timescales. On one hand, Granato
et al.\ (2004) argue that $\tau_{*}$ is small for the most massive protogalaxies, which are then expected to
be the first to form in dramatic star bursts at high redshift. On the other hand Baugh et al.\ (2004b) 
get good fits to current observational data with their superwind semi-analytic model, in which
high constant values of $\tau_{*}$ are associated with quiescent (disk)
star formation so that most of the stellar mass in giant 
ellipticals is built up within disks, and simply rearranged into spheroids during mergers.
Deep HI and continuum\footnotemark surveys with the SKA will determine the true situation.
Further crucial physical information on distant proto-elliptical galaxies will accrue from SKA
HI absorption studies, both of `typical' systems and systems observed during feedback events 
(Kaneker \& Briggs 2004; Jarvis \& Rawlings, 2004).

\footnotetext{
Continuum surveys to an rms depth $\sim 30 ~ \rm nJy$ `come for free' alongside
the HI surveys. Note, however, that scientific exploitation of the continuum
surveys typically needs the full
($\ltsimeq 0.1$ arcsec at 1.4 GHz) angular resolution of the SKA (e.g.\ for the weak
lensing experiment described in Sec.~\ref{sec:exp2}), whereas most of the power of the HI
survey comes from low-resolution data from the `SKA core'. It is plausible that 
the core will be surveying the sky before significant numbers of long baselines are available.
}

The origin of dwarf galaxies will also be open to direct investigation with the SKA. The lack of 
dwarf galaxies relative to dark matter in Fig.~\ref{fig:ps} is normally ascribed, at least in
part, to the suppression of their formation after the epoch of reionization (e.g.
Benson et al.\ 2003). The SKA will have the sensitivity to detect the star-formation
signature of these objects to $z \sim 10$ (V2004) 
and hence can look for direct evidence that the
energy input into baryons during reionization has a profound impact on the formation of these
objects.

\section{Cosmology with the SKA}
\label{sec:cosmology}

\subsection{Context}
\label{sec:context_cosmo}

Probing the nature of dark energy represents the new frontier in cosmology.
The spirit here (and BABR) is to use the measurement of $w$ as an example
of how improving the precision of cosmological parameter
estimation from the $\sim 10$-per-cent level to
the $\sim 1$-per-cent level is likely to yield fundamental advances in 
physics.\footnotemark 
We illustrate this with a series of specific experiments which represent our `best guess'
as to the key techniques to pursue with the SKA.
In Sec.~\ref{sec:wild} we make some more speculative remarks concerning 
other potential cosmological discoveries with the SKA.

\footnotetext{
Some physicists have a firm belief that the dark energy is Einstein's $\Lambda$, and hence that
$w = -1$ at all times and places. We caution that only a decade or so ago many physicists had an equally 
firm belief that $\Omega_{\rm m}=1$ and that dark energy didn't exist at all!
}

\subsection{Why CMB studies are limited}
\label{sec:cmb}

There has been amazing progress in measuring cosmological parameters using observations of the 
CMB, but a few caveats are important to bear in mind. 
Although the WMAP CMB results are extremely impressive, the CMB tells 
us a complicated mixture of information on all of the cosmological 
parameters. Therefore the majority of the key results on individual 
cosmological parameters using the WMAP data have actually come from 
combining it with other information, such as that from galaxy redshift 
surveys (e.g.\ Spergel et al. 2003).
The significance of `orthogonal' constraints in the key slices of parameter space are best illustrated by the 
combination of current CMB constraints (measuring spatial flatness, i.e. $\Omega_{\rm M} + 
\Omega_{\rm DE}$), with independent measurements using
high-redshift supernovae (measuring a very different combination of $\Omega_{\rm DE}$ and $\Omega_{\rm M}$).
This gain over `CMB only' results, 
together with the importance of targeted measurement of key parameters such as 
$H_{0}$ (e.g.\ the HST Cepheid variable Key Project; Freedman et al.\ 2001), is 
emphasised by Greenhill (2004). 
There is now a satisfying agreement 
between constraints from CMB observations, galaxy redshift surveys, supernovae,
cluster abundances and measurements of $H_{0}$ (e.g.\ Spergel et al. 2003).
This is why most cosmologists now believe in a
dark energy component which does not cluster with the galaxies and which is now
causing an acceleration in the expansion of the Universe . 

Another important lesson to
be learnt from current CMB-based studies
is the danger of `hidden theoretical priors'. It is widely believed that CMB 
observations have proven that the Universe is spatially flat, whereas, as emphasised
by Efstathiou (2003), there is not yet adequate precision to rule out a Universe
with very significant spatial curvature; precise spatial flatness remains an assumption made
on the basis of a theoretical belief in cosmic inflation.
Relaxation of this assumption considerably increases uncertainties on
all the other cosmological parameters.

After its launch in 2007, the Planck satellite will provide the next observational step forward in 
CMB astronomy.
However, there are good reasons to believe that it will not {\em on its own} yield great breakthroughs in 
studies of
dark energy. This is because the direct influence of dark energy becomes apparent only in the
older Universe. Its influence on the CMB is subtle, e.g.\ the precise angular
positions of features through the angular-diameter distance $D_{\rm A} (z=1100)$, and
secondary anisotropies like the late-time Integrated Sachs Wolfe (ISW) effect.
Also, error bars on measurements of the CMB temperature power spectrum will only be significantly improved
on relatively small angular scales since existing data on large scales (i.e.\ multipoles $\ell > 200$) 
are already limited by cosmic variance.\footnotemark
CMB temperature data can only be significantly improved in the regime ($\ell \gtsimeq 500$) currently 
covered by small-sky-area, ground-based experiments. 

\footnotetext{
For angular (CMB) measurements the cosmic variance scales as the reciprocal
of $(2 \ell + 1) f_{\rm sky}$, where $f_{\rm sky}$
is the fraction of sky covered.
}

\begin{figure}[htb]
\includegraphics[width=7.0cm,angle=0]{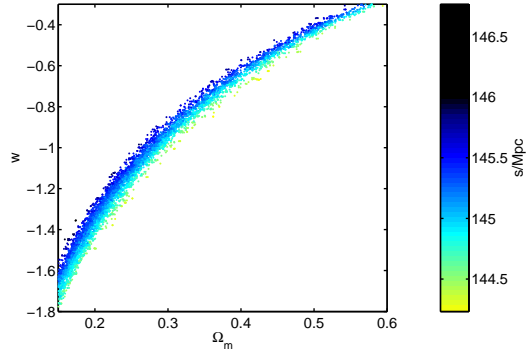} 
\vspace*{-0.3in}
\caption{\small 
\label{fig:sarah1}
Simulated joint constraints on $\Omega_{\rm m}$ and 
$w$ from Planck observations of the CMB (see Fig.\ 1 of BABR for
further details). Points were generated using Monte-Carlo techniques, their
density is proportional to probability and they are colour-coded according to the value of the sound
horizon at recombination $s$ (measured in comoving Mpc).
The strong degeneracy in the $(\Omega_{\rm m},w)$
plane arises because CMB power spectra are approximately invariant
for fixed values of $s  / D_A(z=1100)$ and $\Omega_{\rm m} h^2$ [roughly speaking, the former is held
constant by trading off $\Omega_{\rm m}$ against $w$, and the latter by trading off $\Omega_{\rm m}$
against $h$ so that $h$ decreases systematically from bottom-left to top-right].
The inclusion of an additional tight Gaussian prior
on $h$ would break this main
degeneracy but a {\it secondary degeneracy} between $w$ and $s$
would remain. Without addition information, the `standard ruler' of size
$s$ is `squashable/extendable' at the few-per-cent level (see Sec.~\ref{sec:cmb}) so that
even if $\Omega_{\rm m}$ and $h$ were known precisely, Planck data could not 
determine $w$ to a precision better than $\pm 0.1$.
}
\end{figure}

In Fig.~\ref{fig:sarah1} we show the results from 
simulated measurements with Planck. The value of the combination 
of parameters $\Omega_{\rm m} h^{2}$
will be fixed to 1 per cent accuracy, but the wide range in allowable values of $\Omega_{\rm m}$
results from the well known difficulty, with CMB data alone, of breaking the degeneracy between these parameters.
Let's assume for now that this degeneracy has been broken by an extremely accurate (1 per cent)
measurement of $H_{0}$ (Sec.~\ref{sec:exp3}), meaning $\Omega_{\rm M}$ is then known to few-percent accuracy.
Even then, it is clear from Fig.~\ref{fig:sarah1} that the dark energy parameter $w$ cannot
be measured to any better accuracy than $\pm 0.1$, which is not that far from the accuracy of
current measurements. The reasons for this are easy to understand when we look at the
simulated points in Fig.~\ref{fig:sarah1}, colour-coded according to the physical size 
(in comoving Mpc) of the sound horizon $s$ at recombination. Even assuming exact spatial flatness, 
irreducible uncertainties result from the fact that
this `standard ruler', represented by the value of $s$, is `squashable/extendable' at the few-per-cent level.
It is impossible to tell the difference between an intrinsically smaller (slightly squashed) standard
ruler and a greater angular diameter distance to recombination $D_{A} (z=1100)$ resulting from,
at any $\Omega_{\rm M}$, a more negative value of $w$.

As has already been shown by the improvements obtained by
combining existing CMB data and galaxy redshift surveys of the 
local Universe (e.g.\ the 2dFGRS; Percival et al.\ 2002), the Planck CMB data 
would benefit enormously from being
combined with a very high precision measurement of the galaxy power
spectrum $P(k)$. By measuring redshifts for all galaxies in the sky to $z \sim 1.5$, one
covers a cosmic volume $V \sim 500-$times larger than the 2dFGRS, and it will be
possible to drastically reduce 
error bars on the power spectrum at all wavenumbers $k$. It will also be possible to measure 
the large-scale (small $k$) modes for
which `nuisance' effects such as non-linear growth, peculiar velocity and 
galaxy bias are minimised (Fig.\ 2 of BABR). In Sec.~\ref{sec:exp1}
we argue that an `all hemisphere' HI survey with the SKA would deliver the accurate
measurement of $P(k)$ required for the measurement of $w$.

\subsection{Experiment I: wiggles in the galaxy power spectrum}
\label{sec:exp1}

This experiment (Sec.~3 of BABR) aims to map out the 
acoustic oscillations, or wiggles, in the power spectrum $P(k)$ as a function of redshift.
These wiggles are seen in the CMB temperature power spectrum but 
have yet to be detected in any redshift survey of discrete objects. One way of seeing
why this is such a powerful technique comes from the following crude argument (see e.g.\ Hu \&
Haiman 2003 for a thorough discussion). Imagine rods of fixed (comoving) physical size $s$ 
(the sound horizon at recombination; see Sec.~\ref{sec:cmb}) arranged to be perpendicular to
the line of sight. Measurement of the angular size of these rods, via an SKA redshift survey,
will give a measurement of $s  / D_{\rm A}$ at some effective redshift $z_{\rm eff} \sim 1$. Together
with the measurement of $s / D_{\rm A} (z=1100)$ from the CMB (Sec.~\ref{sec:cmb}), there are 
now two equations which can be combined to cancel $s$, and make accurate measurements of $w$ using the 
ratio $D_{\rm A} (z=1100) / D_{\rm A} (z=z_{\rm eff})$. Extra information comes from rods arranged with
significant radial components since radial lengths are dependent on the $w-$dependent Hubble constant $H(z)$ and,
on large scales, isotropy can be assumed just as in the classic
Alcock-Paczynski test. Once $w$ has been fixed, $s$ can then also be accurately measured yielding, finally,
an accurate standard ruler in cosmology. The redshift range $0 \leq z \leq 1.5$ is key for these 
experiments because
dark energy probably affects $H(z)$ only at low to intermediate $z$, but one must reach a redshift $z \sim 1.5$ 
to both accrue a sufficiently huge cosmic volume and to ensure that 
not too many of the wiggles have been erased by non-linear gravitational clustering
(Fig.\ 4 of BABR). 

This `wiggles' experiment can, of course, be performed by redshift surveys with optical/near-infrared telescopes
and these are being planned (e.g.\ Blake \& Glazebrook 2003). 
The key advantage for at least of some realizations of the 
SKA is the capability of a very wide field of view, making `all hemisphere' surveys a realistic
possibility.\footnotemark 
As, roughly speaking,
error bars on $w$ scale as the square root of the cosmic volume surveyed (noting from 
Fig.~\ref{fig:dndz} that cosmic variance dominates over shot noise due to the
high surface density of HI-emitting galaxies),
only the SKA looks capable of delivering the $\pm 0.01$ accuracy on $w$ required
to make a stringent test of the cosmological constant `model' for dark energy (Fig.\ 5 of
BABR).

\footnotetext{
There are physical limits on the field of view of $\sim 1 ~ \rm deg^{2}$ for
multi-object spectrometers on large optical/near-infrared telescopes, making
`all hemisphere' optical/near-infrared redshift surveys impracticable.
}

AR2004 have demonstrated that `all hemisphere' SKA surveys are needed because one gains
cosmic volume linearly with time as one covers sky area. However, once this survey is 
completed, it will clearly be worth going deeper in redshift because, although the volume gains are
slower, and the power of the tests on $w$ may decline, this will be the only way of beating down cosmic
variance still further. Assuming, as in Sec.~\ref{sec:gformation} $FOV = 10 ~ \rm deg^{2}$, and hence 
a field of view for $z \sim 3$ HI-emitting galaxies of $160 ~ \rm deg^{2}$ (assuming $p=1$), it 
would take 10 years to survey the whole hemisphere to $z \sim 3$. This suggests once more 
(c.f.\ Eqn.~\ref{eqn:fov}) that $FOV = 10 ~ \rm deg^{2}$ is the bare minimum requirement for most cosmological 
studies with the SKA.

What is the competition?
Rather than measuring wiggles and using a `standard ruler'
approach, the proposed JDEM(SNAP) satellite will use high-redshift supernovae as `standardized candles' to make
measurements of the luminosity distance $D_{\rm L}$ as a function of 
redshift (Aldering et al.\ 2004). As $D_{\rm L}$, $D_{\rm A}$ and $H(z)$ are all simply related, this will
produce confidence contours in the
$(w_{0},w_{1})$ plane which are very closely aligned with those of the SKA wiggles experiment. However,
there are worries about whether systematic effects (e.g.\ changes
of properties of supernovae with epoch-dependent galaxy environments) will limit the
SNAP supernova experiment well before the theoretical limits are reached. No serious
systematic problems with the SKA wiggles experiment are expected (BABR).

\subsection{Experiment II: weak gravitational lensing }
\label{sec:exp2}

\begin{figure}[htb]
\includegraphics[width=7.0cm]{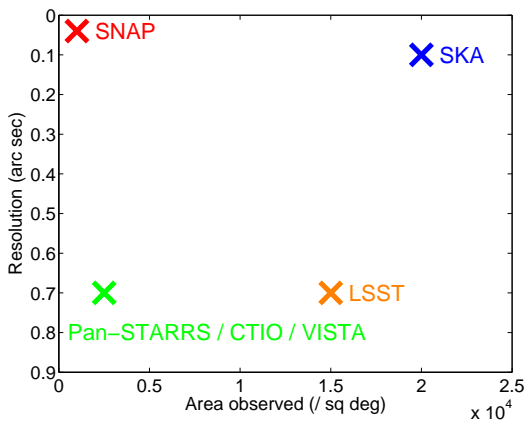} 
\vspace*{-0.3in}
\caption{\small 
\label{fig:sarah2}
The two key properties of upcoming cosmic shear experiments, sky area
surveyed and spatial resolution. Increased sky area reduces the random
uncertainties on the dark energy parameters, whereas increased spatial
resolution reduces the systematic errors. The SKA can excel in {\em both} these directions.}
\end{figure}

We highlighted in Sec.~\ref{sec:cmb} the power of obtaining `orthogonal' constraints on
cosmological parameters, and for dark energy the two parameters often focused upon are
$w_{0}$ and $w_{1}$, the present day value and epoch (redshift) variation of the
dark energy parameter $w$. 
This is also the philosophy behind dark energy experiments with SNAP, hence the SNAP team proposal
of a weak gravitational lensing experiment alongside their supernova experiment
(Refregier et al.\ 2004). This would yield significantly different constraints on
$w_{0}$ and $w_{1}$ because as well as measuring the geometrical effects of 
dark energy, the cosmic evolution of the dark matter power spectrum (as measured
by weak lensing) reflects the competing effects of gravity/dark energy on the growth/dispersion 
of large-scale structure. However, as illustrated by Fig.~\ref{fig:sarah2}, it is the small fields of view
delivered by the SNAP weak lensing surveys that is the critical limitation.
It is worth emphasising here (Fig.\ 6 of BABR) that achieving
very wide fields of view for weak lensing is important not only for beating down 
cosmic-variance-dominated\footnotemark error bars, but also for making precise measurements at scales
much larger than those at which strong, non-linear gravitational clustering effects are important
(and difficult to account for).

\footnotetext{
Cosmic-variance-limited measurements of the dark-matter power spectrum, like those of $P(k)$, 
have error bars scaling as cosmic volume $V^{-0.5}$, and hence as $f_{\rm sky}^{-0.5}$. 
}

Although there are uncertainties in how well SKA surveys will measure cosmic shear (due largely
to the, as yet, unknown shapes of objects in the faint radio population), it
seems very likely (BABR) that there will be a more-than-adequate surface density of 
suitably shaped background sources for measuring shear. The well behaved point spread 
function of a synthesis array like the SKA should ensure superb image quality, 
and the extra leverage supplied by `lensing tomography' (Hu 1999)
should accrue from the redshift information available from HI surveys.

\begin{figure}[htb]
\includegraphics[width=5.0cm,angle=-90]{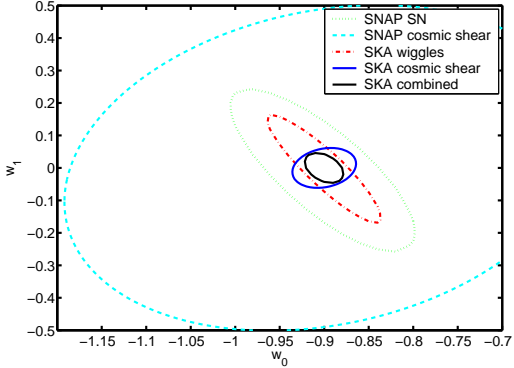} 
\vspace*{-0.3in}
\caption{\small 
\label{fig:answer}
Comparison of constraints on $w(z) = w_0 + w_1 z$ obtained by
the proposed satellite mission JDEM(SNAP), using both supernova and
cosmic shear experiments, and the SKA, using acoustic oscillations and
cosmic shear. The contours represent `$\pm 1 \sigma$' confidence intervals.
A Gaussian prior upon the matter density,
$\sigma(\Omega_{\rm m}) = 0.01$, was consistently applied to the
supernova and acoustic peaks analyses.  For the latter we also assume
a prior $\sigma(h) = 0.01$ for the Hubble parameter (Sec.~\ref{sec:exp3}).  
We assumed that the SNAP supernova experiment measures the magnitudes
of 2000 supernovae evenly spaced between $z = 0.1$ and $z = 1.7$ with an
uncertainty of $0.15$ per supernova; we combined this with the 300
supernovae at $z=0.05$ expected from the `SNFactory' (Aldering et al.\ 2002) and marginalised over the 
intrinsic magnitude uncertainty.
The cosmic shear contours for both SNAP and SKA are
conservative because the sample has not been split into two or more
bands in redshift. This analysis method was shown by Refregier et
al.\ (2004) to improve significantly the constraints from SNAP cosmic
shear, and would have a similar effect on the SKA measurements.}
\end{figure}

Cosmic shear experiments likely to happen before those performed by the SKA 
are represented in Fig.~\ref{fig:sarah2}. An increased sky area is essential for small
random errors on the dark energy parameters, hence the plans for the
large optical telescope (the LSST; http://www.lsst.org/lsst\_home.shtml) 
which will image large sky areas optically. However it is unknown at
what point ground based optical experiments will be hampered by
systematic effects associated with the measurement of
galaxy shears using ground-based (seeing-limited) optical data.
This is of course why the SNAP (space-based) weak lensing experiment is judged worthwhile, but note here
how, assuming a wide field of view realization for the SKA, SNAP efforts on lensing may be quickly trumped
by surveys with the SKA. 
The improvements on measurements of dark energy with the SKA are illustrated in Fig.~\ref{fig:answer}.

\subsection{Experiment III: $H_{0}$ from water masers}
\label{sec:exp3}

In the early 21$^{\rm st}$ century, the combination of CMB (Planck) and galaxy redshift
survey (SKA) data will make measurements of $w$ and its time variation with unprecedented 
levels of accuracy (see BABR). The `nuisance parameter' $h$ will be measured along the way, but
learning the lessons of Sec.~\ref{sec:cmb}, the conclusions concerning dark energy will be particularly 
compelling if $h$ is also measured independently to high (1-per-cent) accuracy. 
Current 10-per-cent-accuracy estimates
of $H_{0}$ are obtained via measurements of `Standard Candles' that are limited by
intrinsic scatter and systematic errors.  Even the best estimate is
calibrated using the controversial distance to a single and possibly
unrepresentative (metal-poor) galaxy, the LMC. The SKA is likely to play a central r\^{o}le in 
improving on this situation (Greenhill, 2004). 

Moving to a 1-per-cent level of accuracy for the measurement of $H_{0}$ will 
require a new method which, to overcome systematics, measures reliable {\em geometric} distances to galaxies, 
and which, to overcome peculiarities in the local Hubble expansion rate, reaches to
reasonable redshifts. 
This can be achieved by using the SKA to find, map, and
monitor water maser sources\footnotemark
\footnotetext{The rest-frame frequency of the water maser line is $\approx 22 ~ \rm GHz$, so this
requires high-frequency capability for the SKA.} 
within (roughly edge on) accretion disks at
sub-parsec radii (from supermassive black holes) out to distances of a few hundred Mpc. Although only a
fraction of masers turn out to be relatively simple dynamical systems
suited to detailed modelling, a very large number of masers will be
discovered and the absolute number of suitable systems will be orders of
magnitude greater than today. The result will be systems with well measured 
black hole masses, disk geometries, orientations, and mappings between 
angular and physical sizes. 

Amongst the compelling features of this method are: (i) evidence from the remarkable study of NGC4258 
(Herrnstein et al.\ 1999; Humphreys et al. 2004) that the method works
well in practice; (ii) relatively few
sources of systematic uncertainty given highly constrained models for the most suitable systems;
and (iii) the expectation of uncorrelated uncertainties in 
individual distance measurements that should ensure that accuracy on $H_{0}$ increases 
as $\sqrt{N}$, where, to get to the required 1-per-cent level of accuracy, $N$ need only be a few hundred 
out of the many thousands of extragalactic water maser systems that will be discovered by the SKA.

\subsection{Cosmology beyond dark energy}
\label{sec:wild}

We have demonstrated that the exquisite galaxy power spectrum $P(k)$ 
measured by the SKA will, together with data from Planck and other CMB experiments, provide the basis for 
major advances in cosmology. We have used studies of the dark energy parameter $w$ to illustrate
this. Here, we assume instead that, with much improved accuracy on measuring $w$, dark energy still
appears to behave like Einstein's cosmological constant. In this case, what would the SKA
generate in the way of important new results? Here, in rough order of the wildness of the speculation, 
are a few possibilities.

\begin{itemize}

\item Can we detect the effects of dark energy through the late-time ISW effect?
Gravitational potential wells decay rather than grow under the influence of dark energy, 
and there have been claims of a detection of the predicted correlation between
large-scale ($\ell \ltsimeq 10$) CMB anisotropies and the low-redshift matter distribution
(e.g.\ Boughn \& Crittenden 2002). The combination of Planck and SKA data will establish the 
reality of the ISW signal, and can be used to establish the clustering properties of dark energy
(e.g.\ Weller \& Lewis 2003).

\item Can we tie down the physics of inflation? Combining CMB and SKA
measurements will allow new measurements of the scalar spectral index of the power spectrum
of fluctuations. The value of this parameter, if it is indeed a single parameter, or, as is more likely, its
function of scale (the so-called running scalar spectral index) are predicted outputs from specific 
inflationary models. The CMB power spectrum is also influenced by tensor (gravitational wave) modes
whereas the SKA $P(k)$ is not. Analysis of combined CMB/SKA datasets will have great power in measuring the 
running scalar spectral index and isolating the tensor component in the CMB.

\item Are the primordial fluctuations purely adiabatic?
It is commonly assumed that the primordial perturbations are in energy density (adiabatic), but more
general models add in perturbations in entropy density (isocurvature) which, being spatially
homogeneous in energy density, leave no perturbation in spatial curvature. Bucher et al.\ (2004) have shown that the
existence of such modes cannot be ruled out by current datasets, implying that the 
power of the combined CMB/SKA dataset will be needed to make significant progress. Again, such
measurements allow sensitive tests of models for inflation.

\item Are the primordial fluctuations Gaussian? Searches for evidence
of non-Gaussianity in the statistics of the primordial fluctuation spectrum 
have, from CMB data alone, so far proved negative (Komatsu et al.\ 2003). The
SKA `all hemisphere' survey will detect large numbers of superclusters, the
evolved counterparts of rare fluctuations in the quasi-linear regime, with which to make
sensitive independent tests.

\item
Do the values of the physical constants change with time?
As reviewed by Curran et al.\ (2004), there have been recent claims that 
the value of the fine structure constant $\alpha$ varies with cosmic epoch. Curran et al.\ demonstrate that 
SKA observations of redshifted radio absorption lines hold the key to confirming this remarkable
result and to discovering changes in other physical constant (e.g.\ the ratio of electron and
proton masses). Many theories for unifying gravity with other fundamental forces predict such
variations, and the SKA may measure them.

\item Are there sharp features in $P(k)$ and are they a signature of `transPlanckian' physics?
The binned power spectrum $P(k)$ derived from CMB results has some `glitches'
that have been interpreted as the results of new physics on scales smaller than the 
Planck length which were imprinted on $P(k)$ during inflation (e.g.\ Martin \& Ringeval 2004).
Precision large-scale measurements of $P(k)$ with the SKA will establish the reality and
location of any glitches, and look for evidence of such new physics.

\item Is the Universe precisely spatially flat? Recalling the discussion of hidden theoretical
priors in Sec.~\ref{sec:cmb}, we note that one explanation of the low quadrapole signal 
in the CMB is that the Universe has a small, but positive, curvature and a $P(k)$ which cuts 
off on this curvature scale (e.g.\ Efstathiou 2003). Precision cosmology led by the SKA could, 
for example, confirm that $w = -1$, and that the Universe is closed.

\item Can we see the copies of our own Local Group predicted in a small Universe? 
Another explanation of the low CMB quadrapole is that the Universe is small but 
has a weird topology, being small but multiply connected (e.g.\
Cornish et al.\ 2004). If the length scale of its fundamental domain is smaller than 
the particle horizon, then it should be possible to see a `ghost' of our local environment at an earlier
epoch. A great triumph of a semi-analytic model for galaxy formation would be a
prediction of the past properties of the Local Group that was so precise that a 
recognisable ghost pattern of the Local Group could be seen at
high redshift. With its huge sensitivity (e.g.\ to the young, starbursting precursors of
the Milky Way and Andromeda) and sky coverage, the SKA would be the
premier instrument to search for such effects.

\end{itemize}

\section{Concluding remarks}
\label{sec:conclusions}

Studies of galaxy evolution and dark energy can be revolutionised by the SKA but are 
demanding on the design parameters of its realization. 

The HI surveys are obviously 
a `unique selling point' for the SKA, and through the necessity of detecting 
galaxies out to at least redshift $z \sim 3$, provide a simple justification for
the raw sensitivity needed. We have argued that cosmological applications, illustrated
mainly here by the acoustic oscillations (`wiggles') experiment, requires
the ability to do `all hemisphere' surveys mapping $\sim 10^{9}$ galaxies
to $z \sim 1.5$, and hence wide instantaneous fields of view (Eqn~\ref{eqn:fov}).

Continuum surveys are vital for galaxy evolution studies as they
can detect star-forming galaxies of all types out to extreme redshifts. In cosmology, weak
gravitational lensing experiments with these continuum surveys are likely to be revolutionary given the 
unique SKA combination of wide field of view and stable
point spread function. 

We have also argued that a high-frequency capability\footnotemark for the SKA may be vital for
cosmological studies because of the great promise shown by extragalactic water masers as the future 
method of choice for measuring $H_{0}$ to 1-per-cent accuracy.

\footnotetext{
Experiment III (Sec.~\ref{sec:exp3}; Greenhill 2004) requires the SKA
frequency range reaches $\approx 22$ GHz to observe water masers, and requires
long (intercontinental) baselines with high spectral-line sensitivity.
}

This combination of experiments proposed with the SKA will be essential in transitioning from the
10-per-cent accuracy currently achieved by `precision' cosmology to the 
1-per-cent accuracy needed, for example, to make the first precision study of the properties of dark energy.
The SKA datasets will revolutionise studies of galaxy formation and evolution
by measuring the way in which galaxies trace dark-matter structures (i.e.\ bias) and by determining the
physical processes that generated all types of galaxies from low-luminosity dwarves to the
most massive ellipticals. 

In galaxy evolution studies, the SKA will add unique information from the radio band to data to be obtained 
by other upcoming facilities like ALMA, JWST, giant optical telescopes and high-energy
satellite missions. In cosmological studies, a 
wide-field-of-view realization of the SKA can easily outperform projects like LSST and JDEM(SNAP) 
which are largely
being designed to answer questions concerned with dark energy. The SKA will
replace optical telescopes as the machine for delivering the huge redshift surveys needed to
push forward cosmological research beyond what is possible with CMB-dominated datasets.
The SKA should aim, like the Planck satellite for CMB studies,
to provide results limited by cosmic variance rather than sky area coverage.

\section*{ACKNOWLEDGEMENTS} 
SR is grateful to the UK PPARC for a 
Senior Research Fellowship, and for financial support from the Australia 
Telescope National Facility. 
FBA thanks PPARC for a Gemini Research Studentship and the SKA Project Office for
financial support.  
SB and CMB are supported by the Royal Society.
CB acknowledges Karl Glazebrook for a valuable collaboration
developing simulations of acoustic oscillations and also 
the SKA Project Office for financial support. We thank Chris Carilli, Jo Dunkley,
Phil Marshall and Richard Massey for useful discussions.

\end{document}